\documentclass[aip,pop,10p,sanserif,reprint,twocolumn]{revtex4-1}

\usepackage[english]{babel}
\usepackage{amsmath, amsthm, amssymb, latexsym, graphicx}
\usepackage[utf8]{inputenc}
\usepackage{color} % Font Color 
\newcommand{\eps}{\varepsilon}
\newcommand{\del}{\delta}
\newcommand{\om}{\omega}
\newcommand{\na}{\boldsymbol\nabla}

\begin{document}

\title{Effects of electron inertia in collisionless magnetic reconnection}

\author{Nahuel Andr\'es}
\email[E-mail: ]{nandres@iafe.uba.ar. Tel.: +5411 47890179 (int. 134).}
\affiliation{Instituto de Astronom\'ia y F\'isica del Espacio (CONICET-UBA), Argentina}
\affiliation{Departamento de F\'isica (FCEN, UBA), Argentina}

\author{Luis Martin}
\affiliation{Departamento de F\'isica (FCEN, UBA), Argentina}

\author{Pablo Dmitruk}
\affiliation{Departamento de F\'isica (FCEN, UBA), Argentina}

\author{Daniel G\'omez}
\affiliation{Instituto de Astronom\'ia y F\'isica del Espacio (CONICET-UBA), Argentina}
\affiliation{Departamento de F\'isica (FCEN, UBA), Argentina}

\date{\today}

\begin{abstract}
We present a study of collisionless magnetic reconnection within the framework of full two-fluid MHD for a completely ionized hydrogen plasma, retaining the effects of the Hall current, electron pressure and electron inertia. We performed 2.5D simulations using a pseudo-spectral code with no dissipative effects. We check that the ideal invariants of the problem are conserved down to round-off errors. Our numerical results confirm that the change in the topology of the magnetic field lines is exclusively due to the presence of electron inertia. The computed reconnection rates remain a fair fraction of the Alfv\'en velocity, which therefore qualifies as fast reconnection.
\end{abstract}


\maketitle

\section{Introduction}\label{intro}

Magnetic reconnection is an important physical mechanism of energy conversion in various space plasma physics environments, such as the solar corona or planetary magnetospheres\citep{Bi2000,PF2007}. This process locally changes the magnetic field topology, transforming free magnetic energy into kinetic energy and heating of the plasma. To study the efficiency of magnetic reconnection, the reconnection rate is considered. Theoretical models of magnetic reconnection were first developed within the framework of one-fluid resistive magnetohydrodynamics (MHD), the so-called Sweet-Parker regime \citep{S1958,P1957}. In the Sweet-Parker model, the reconnection rate scales as the square root of the magnetic resistivity, which leads to exceedingly low reconnection rates for most space physics environments \citep{E1981,I2007,E2007,B2004,Y2011}. A possible solution to this problem was reported by \citet{P1964}, giving rise to  the concept of fast reconnection. However, numerical simulations showed that the classical 
Petschek model configuration cannot be attained in a model with a spatially uniform resistivity \citep{Bi1986}.

The idea that MHD turbulence may play an important role in a magnetic reconnection setup was first proposed by \citet{M1986}, by adding turbulent fluctuations on a two-dimensional sheet pinch configuration. For a specific model for MHD turbulence \citep{GS1995}, \citet{L1999} reported that the rate of magnetic reconnection is increased in the presence of a stochastic or turbulent component of the magnetic field. In their model, the fast reconnection speed is determined by the level of large-scale kinetic energy feeding the turbulent cascade, which was confirmed by \citet{K2009} using direct numerical simulations. Within the framework of resistive MHD, \citet{B2009} showed that thin current sheets with Lundquist number exceeding a critical value, are unstable to a super-Alfv\'enic tearing instability. As a result of this instability, the system reaches a nonlinear reconnection rate which is larger than the Sweet-Parker rate by an order of magnitude. Recently, \citet{Y2010} reported an extensive review on 
magnetic reconnection, discussing results from theory, numerical simulations, observations from space satellites, and recent results from laboratory plasma experiments. 

Kinetic plasma effects such as Hall and electron inertia, introduce new spatial and temporal scales into the theoretical fluid description. At length scales larger than the so-called \textit{ion skin-depth}, these two effects can be neglected. For instance, if the resistive scale is larger than the ion skin-depth the resistive MHD model is a valid description for a collisional plasma. On the other hand, at scales below the ion skin-depth, the Hall-MHD (HMHD) description is valid. In this scenario, the ions are no longer frozen-in to the magnetic field lines as a result of the Hall current term. Meanwhile, the electrons remain frozen-in to the magnetic field lines. \citet{Sm2004} examined the influence of the Hall effect and level of MHD turbulence on the reconnection rate in 2.5D compressible Hall MHD. Their results indicate that the reconnection rate is enhanced both by increasing the Hall parameter and the turbulence amplitude. In any of the cases discussed above, a small amount of magnetic resistivity is 
necessary to break the frozen-in condition and start the reconnection process.

\citet{Bi1997} reported theoretical studies of collisionless magnetic reconnection within the framework of two-fluid theory. In particular, the authors propose that reconnection is controlled by the whistler mode, leading to the decoupling of ions from electrons on scales of the order of the ion skin-depth, where the behavior of the plasma is approximately described by the equations of electron-MHD (EMHD).  In this approximation, which becomes asymptotically valid at spatial scales smaller than the ion skin-depth, ions are considered static (because of their larger mass) and electrons are the only species to carry the electric current. {More recently, \citet{Z2008} considered the potential relevance of electron viscosity on the reconnection rate, which is also known as hyper-resistivity, since the effect is represented by a $\nabla^4$ term in the induction equation. Using scaling arguments on a steady state configuration, they find that the hyper-resistive regime can potentially lead to fast 
reconnection, even though the length of the electron diffusion region might depend explicitly on the level of hyper-resistivity. \citet{S2009} also study the role of hyper-resistivity in the somewhat more general framework of HMHD performing two dimensional simulations, and confirm the previous results. Electron viscosity corresponds to a particular closure on the electron pressure tensor within the fluidistic description. Considering alternative closure approximations on the electron pressure tensor,  \citet{C2009} performed a linear analysis on the EMHD equations to study the role of electron-pressure anisotropies on the evolution of the tearing-mode instability. They find that the relative importance between electron-pressure and electron inertia effects during reconnection, depends on the ratio between the thermal electron Larmor radius and the electron skin depth. \citet{H2011} report a comprehensive study of anisotropies of the electron pressure tensor on the reconnection process, in the cases of 
either presence or absence of a guide magnetic field.}

Geospace Environment Modeling (GEM) Reconnection Challenge \citep{Bi2001} was a project designed to study collisionless magnetic reconnection assuming different theoretical approaches. Using fully electromagnetic particle in cell \citep{H2001,P2001,S2001}, resistive MHD, HMHD \citep{BH2001,O2001,Ma2001,S2001} and hybrid codes \citep{K2001,S2001}, the authors studied a simple 2D Harris current sheet configuration with a specified set of initial conditions. They found that the reconnection rate is insensitive to the mechanism that breaks the frozen-in condition, and corresponds to an inflow velocity of nearly 10 \% of the Alfv\'en speed. In addition to these arguments, \citet{S1999} claimed that the reconnection rate is found to be a universal constant as the system become very large. However, several studies have demonstrated that the reconnection rate might still depend on the value of the Hall parameter \citep{S2008,Bi1997,Sm2004,Mob2005,Mo2006} or on the level of turbulent fluctuations \citep{L1999,Sm2004}.
 Moving beyond the steady-state models, \citet{OP1993} showed that electron inertia can lead to growth rates faster than exponential in time. This work was made under the assumption that the nonlocal ion motion can be neglected. \citet{C2013} reported results including ion gyration effects. These authors have shown analytical evidence that the qualitative differences between hot and cold ion reconnection is linked to the formation of strong electric fields due to ion gyration effects. {Recent particle-in-cell (PIC) simulations of collisionless magnetic reconnection studied the effect of both electron inertia and non-gyrotropic (off-diagonal) pressure tensor effects \citep{H1999}. Using 3D electromagnetic PIC simulations, \citet{F2011} confirmed the presence of a fast reconnection rate. For configurations displaying translational symmetry along the current sheet, he also finds that the electrons cross the diffusion region without thermalization, which means that the magnetic dissipation is dominated by 
electron inertia (see also  \citet{F2009}).  Using 2D PIC simulations, \citet{Z2011} found that the size of the central dissipation region is controlled by the electron to ion mass ratio, even though the reconnection rate is largely insensitive to this mass ratio.}

In a two-fluid description of a plasma, at least two kinetic effects are able to break magnetic field lines and give rise to reconnection: electric resistivity and electron inertia. {\citet{Al2003} considered the relative importance between these two effects on externally driven magnetic reconnection. They find that when the boundaries are perturbed at rates slower than the hydrodynamic time and faster than the resistive time, a current sheet as narrow as the electron skin depth forms which undergoes resistive dissipation at later times. } Most if not all of the fluid descriptions listed above include electric resistivity, at the very least a numerical resistivity originated in the computational scheme used to calculate the spatial derivatives. {However, in truly collisionless regimes (i.e. where the collisional frequencies remain much smaller than all other relevant frequencies, such as those corresponding to the cyclotron motions), magnetic reconnection should be driven solely by electron 
inertia.} This physical or numerical resistivity is likely to be the ultimate cause of the reconnection process. In the present paper, our goal is to study magnetic reconnection exclusively due to electron inertia, by completely suppressing the action of electric resistivity. We use a pseudo-spectral scheme to compute the spatial derivatives, which converges exponentially fast as the number of grid points is increased. As a result, we can run simulations with zero resistivity and/or viscosity, and check that we are not spuriously adding numerical resistivity simply by monitoring the energy conservation for each run. We find that energy is conserved with a precision consistent with round-off errors. Therefore, we are certain that reconnection in our simulations arises exclusively as a result of finite electron inertia, and not because of the presence of physical or numerical resistivity. For spatial scales below of the \textit{electron skin-depth} the terms of electron inertia are dominant, and the electrons 
can no longer be frozen-in to the magnetic field lines \citep{V1975}. Only at this level of description, a change in the topology of the magnetic field lines exclusively due to electron inertia (i.e. including the mass of the electron explicitly) becomes possible. We call Electron Inertia Hall-MHD (EIHMHD) to a theoretical framework that extends HMHD and includes the inertia of electrons. This level of description should not be confused with the EMHD approximation \citep{B1992}, for which the ion motion is neglected. Instead, we retain the whole dynamics of both the electron and ion flows throughout all the relevant spatial scales.

In summary, our main goal in this paper is to study the magnetic reconnection process, using a full two-fluid model for a completely ionized hydrogen plasma, retaining the Hall current and electron inertia. To the extent of our knowledge, this is the first time such a complete ideal two-fluid model is presented, considering also a pseudo-spectral method to accurately run simulations with negligible numerical resistivity.

In section \ref{eih-mhd} we develop the EIHMHD model used in the present paper and present the ideal invariants of the model. In section \ref{2.5d} we show the set of equations that describe the dynamical evolution of the problem in a 2.5D setup. The linear modes of this incompressible model, and the numerical code used to integrate the equations are described in section \ref{idealinv}. In section \ref{results} we present our main results and, finally, in section \ref{conclus} we summarize our conclusions.

\section{Electron Inertia Hall-MHD model}\label{eih-mhd}

The equations of motion for a plasma made of protons and electrons with mass $m_{p,e}$, charge $\pm e$, density $n_{p}=n_{e}=n_{o}$ (because of quasi-neutrality), pressure $p_{p,e}$ and velocity $\textbf{u}_{p,e}$ respectively, in the ideal limit can be written as
\begin{eqnarray}\label{motionp}
 m_pn_o\frac{d \textbf{u}_p}{dt} &=&  en_o(\textbf{E}+\frac{1}{c}\textbf{u}_p\times\textbf{B})-\boldsymbol\nabla p_p \\ \label{motione}
 m_en_o\frac{d \textbf{u}_e}{dt} &=&  -en_o(\textbf{E}+\frac{1}{c}\textbf{u}_e\times\textbf{B})-\boldsymbol\nabla p_e \\ \label{ampere}
 \textbf{j}&=&\frac{c}{4\pi}\boldsymbol\nabla\times\textbf{B}={en_o}(\textbf{u}_p-\textbf{u}_e)
\end{eqnarray}
where \eqref{ampere} corresponds to Ampere's law neglecting the displacement current, $c$ is the speed of light and the total derivative is
\begin{equation}
 \frac{d\textbf{u}_{p,e}}{dt} \equiv \frac{\partial\textbf{u}_{p,e}}{\partial t} + (\textbf{u}_{p,e}\cdot\boldsymbol\nabla)\textbf{u}_{p,e}
\end{equation}
The conservation of mass for each species implies
\begin{eqnarray}\label{mass}
 \frac{\partial(m_{p,e}n_o)}{\partial t} + \boldsymbol\nabla\cdot(m_{p,e}n_o\textbf{u}_{p,e}) &=& 0
\end{eqnarray}
This set of equations can be written in a dimensionless form in terms of a typical length scale $L_0$, a constant particle density $n_0$, a value for the magnetic field $B_0$ and a typical value of velocity $u_0=v_A=B_0/\sqrt{4\pi n_0M}$ (the Alfv\'en velocity) where $M\equiv m_p+m_e$,
\begin{eqnarray}\label{dlessp}
 (1-\del)\frac{d \textbf{u}_p}{dt} &=&  \frac{1}{\eps}(\textbf{E}+\textbf{u}_p\times\textbf{B})-\boldsymbol\nabla p_p \\ \label{dlesse}
 \del\frac{d \textbf{u}_e}{dt} &=&  -\frac{1}{\eps}(\textbf{E}+\textbf{u}_e\times\textbf{B})-\boldsymbol\nabla p_e \\ \label{dlesso}
 \eps\textbf{j}&=&\textbf{u}_p-\textbf{u}_e 
\end{eqnarray} 
where we have introduced the parameters $\del\equiv m_e/M$ and $\eps\equiv c/\om_{M}L_0$, and $\om_{M}=\sqrt{4\pi e^2n_0/M}$ is related to the plasma proton frequency $\om_{pp}=\sqrt{4\pi e^2n_0/m_p}$ as $\om_{M}=\om_{pp}\sqrt{m_p/M}$. It is important to mention that in the limit of electron inertia equal to zero, we obtain $\om_{M}=\om_{pp}$, and therefore $\eps=\eps_H=c/\om_{pp}L_0$ which is the usual Hall parameter.

Using the definition of the hydrodynamic velocity field 
\begin{eqnarray}\label{hdvelocity}
 \textbf{u}&\equiv&\frac{m_p\textbf{u}_p+m_e\textbf{u}_e}{m_p+m_e}=(1-\del)\textbf{u}_p+\del\textbf{u}_e
\end{eqnarray}
we can readily obtain the relations between the hydrodynamic variables and the velocity of each species as
\begin{eqnarray}
 \textbf{u}_p&=&\textbf{u} + \del\eps\textbf{j} \\
 \textbf{u}_e&=&\textbf{u} - (1-\del)\eps\textbf{j}
\end{eqnarray}
The modified Euler equation, is the sum of the corresponding equations of motion \eqref{dlessp} and \eqref{dlesse},
\begin{align}\label{meuler}
\frac{d \textbf{u}}{dt} &= \textbf{j}\times\left[\textbf{B}-\del(1-\del)\eps^2\nabla^2\textbf{B}\right] - \beta\na p
\end{align}
where $p\equiv p_p+p_e$ is the hydrodynamic pressure, and $\beta$ is the ratio between the gas pressure and the magnetic pressure. Note that in the limit of negligible electron inertia (i.e., for $\delta\rightarrow0$), equation \eqref{meuler} reduces to the standard equation of motion of one-fluid MHD, and this is also the case for the HMHD description, which is a two-fluid theoretical description, but considering massless electrons. The equation of motion for electrons \eqref{dlesse}, using $\textbf{E}=-\partial_t\textbf{A}-\na\phi$ and $((\textbf{u}_e\cdot\na)\textbf{u}_e)=\boldsymbol\om_e\times\textbf{u}_e+\na(u_e^2/2)$ can be cast into 
\begin{eqnarray}\label{mote}
\frac{\partial}{\partial t}(\textbf{A}-\del\eps\textbf{u}_e) &=& \textbf{u}_e\times(\textbf{B}-\del\eps\boldsymbol\om_e)+ \nonumber \\
&+&\na(\eps p_e+\del\eps\frac{u_e^2}{2}-\phi)
\end{eqnarray}
We define,
\begin{eqnarray}
\textbf{B}'\equiv \textbf{B}-\del\eps\boldsymbol\om_e&=&\textbf{B}-(1-\del)\del\eps^2\nabla^2\textbf{B} -\del\eps\boldsymbol\om
\end{eqnarray}
where $\boldsymbol\omega=\boldsymbol\nabla\times\textbf{u}$ is the hydrodynamic vorticity. Taking the curl of equation \eqref{mote}, it is possible to obtain a dynamical equation for the magnetic field 
\begin{equation}\label{dynB}
\partial_t~\textbf{B}'= \boldsymbol\nabla\times\left\{\left[\textbf{u}-(1-\delta)\eps\textbf{j}\right]\times\textbf{B}'\right\}
\end{equation}
Again, it is straightforward to verify that for $\delta\rightarrow0$, equation \eqref{dynB} reduces to the induction equation for HMHD.

Just as for three-dimensional Hall-MHD, the Electron Inertia Hall-MHD model has three ideal invariants. Using $\textbf{E}=-\frac{1}{c}\partial_t\textbf{A}-\boldsymbol\nabla\phi$, we can readily show that the total energy $E$ is one of these ideal invariants, where
\begin{equation}\label{general_ene}
 E = \int d^3r \bigg(\sum_sm_sn_s\frac{u^2_s}{2}+\frac{B^2}{8\pi}\bigg)
\end{equation}
The other two ideal invariants are one helicity per species, i.e.
\begin{equation}\label{general_hel}
H_s = \int d^3r \left(\textbf{A}+\frac{cm_s}{q_s}\textbf{u}_s\right)\cdot\left(\textbf{B}+\frac{cm_s}{q_s}\boldsymbol\omega_s\right)
\end{equation}
where $\boldsymbol\omega_s=\boldsymbol\nabla\times\textbf{u}_s$ and in this case $s=p,e$. It is worth to mention that in the Hall-MHD limit, i.e. $\del\rightarrow0$, the conservation of the ion helicity and electron helicity corresponds to the conservation of the hybrid helicity and magnetic helicity respectively \citep{G2008}.

\section{2.5D Setup}\label{2.5d}

In a 2.5D setup, the vector fields depend on two coordinates, say \textit{x} and \textit{y}, although they have three components. Considering the incompressible case, i.e. $\boldmath\nabla\cdot\textbf{u}=0$, we can write the magnetic and velocity fields as
\begin{eqnarray}\label{Bmag}
\textbf B &=& \boldsymbol\nabla\times[\hat{\textbf z}~a(x,y,t)] + \hat{\textbf z}~b(x,y,t)\\ 
\textbf u &=& \boldsymbol\nabla\times[\hat{\textbf z}~\varphi(x,y,t)] + \hat{\textbf z}~u(x,y,t) 
\end{eqnarray}
where $a(x,y,t)$ and $\varphi(x,y,t)$ are the scalar potential for the magnetic and velocity fields respectively. In terms of these scalar potentials, equations \eqref{meuler} and \eqref{dynB} take the form
\begin{eqnarray}\label{1.1}
\partial_t~\omega &=& [\varphi,\omega] - [a,j] - (1-\del)\del\eps^2[b,\nabla^2b]  \\ \label{1.2}
\partial_t~u &=& [\varphi,u] - [a,b] - (1-\del)\del\eps^2[j,b] \\ \label{1.3}
\partial_t~a' &=& [\varphi - (1-\del)\eps b,a']  \\ \label{1.4}
\partial_t~b' &=& [\varphi - (1-\del)\eps b,b'] + [u - (1-\del)\eps j,a'] 
\end{eqnarray}
where 
\begin{eqnarray}
\omega&=&-\nabla^2\varphi \\
j&=&-\nabla^2a \\
a'&=&a+(1-\del)\del\eps^2j-\del\eps u \\
b'&=&b-(1-\del)\del\eps^2\nabla^2b-\del\eps\om
\end{eqnarray}
and the nonlinear terms are the standard Poisson brackets, i.e. $[p,q]=\partial_xp\partial_yq-\partial_yp\partial_xq$. The set of equations \eqref{1.1} - \eqref{1.4} describe the dynamical evolution of the magnetic and velocity fields for the reconnection problem. When $\del=0$ (massless electrons) this set of equations reduces to the incompressible 2.5D HMHD equations \citep{G2008}.

\section{Ideal invariants and linear modes}\label{idealinv}

Linearising equations \eqref{1.1}-\eqref{1.4} around a static equilibrium given by a homogeneous magnetic field of intensity $\text{B}_0$ in the $x$-$y$ plane, we obtain the following dispersion relationship:
\begin{equation}\label{modes}
\big\{ \sigma^2 \big[1+(1-\del)\del\eps^2k^2\big]-k^2\cos^2(\theta_{kB})\big\}^2=\sigma^2\eps^2k^2(2\del-1)
\end{equation}
where $\theta_{kB}$ is the angle between the propagation vector and the equilibrium magnetic field and $\sigma$ is the temporal frequency. The solution of equation \eqref{modes} yields the normal modes of oscillation of equations \eqref{1.1}-\eqref{1.4}.

\begin{figure}
\centering
\includegraphics[width=.41\textwidth]{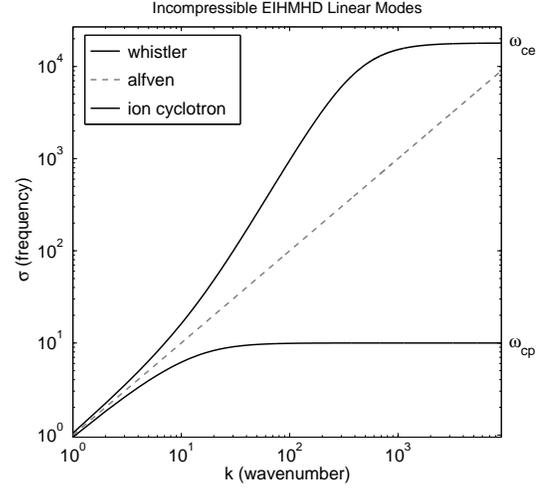}
\caption{Linear propagation modes in EIHMHD$~$ model for a realistic mass ratio, $\theta_{kB}=0$ and $\eps_{H}=0.1$. The dotted line corresponds to the MHD Alfv\'en mode, for reference.}
\label{rel-disp}
\end{figure}

Figure \ref{rel-disp} shows the two modes of propagation of waves in EIHMHD, for a realistic mass ratio of $m_e/m_p = 1/1836$, $\theta_{kB}=0$ and $\eps_{H}=0.1$. The dotted line corresponds to the MHD Alfv\'en mode, for reference. As in HMHD \cite{G2008} the bottom branch represents the shear ion-cyclotron waves, which converges to the proton cyclotron frequency ($\om_{cp}=e\text{B}_0/m_pc$). The top branch corresponds to the whistler branch and, in contrast to HMHD, it reaches a maximum given by the electron cyclotron frequency ($\om_{ce}=e\text{B}_0/m_ec$). The fact that both linear modes have upper boundaries for their frequencies represents an advantage from the numerical point of view, with respect to the unbounded dispersion relation in HMHD. The maximum frequency in EIHMHD (corresponding to $\om_{ce}$) is suggestive of the existence of a minimum time-step in the numerical integration scheme which is independent of the spatial resolution $\Delta t << 1/\om_{max}=2/\om_
 {ce}$. Instead, in HMHD, the whistler branch implies a $k$-dependent maximum frequency $\om_{max} \sim k_{max}^2$ and therefore the minimum time-step in the numerical integration scheme (CFL condition\citep{CFL1928}) depends quadratically on the spatial resolution, $\Delta t = 1/\om_{max} \sim 1/k_{max}^2 \sim \Delta x ^2$. As a result, HMHD is computationally more demanding as the spatial resolution is increased, as compared with the more complete EIHMHD model.

In a 2.5D setup, the dimensionless expressions for the three ideal invariants are
\begin{eqnarray}\label{energy}
 E &=& \frac{1}{2}\int d^2r\bigg[|\boldsymbol\nabla\varphi|^2 + u^2 + |\boldsymbol\nabla a|^2 + b^2 +\nonumber \\
  && (1-\del)\del\eps^2|\boldsymbol\nabla b|^2 + (1-\del)\del\eps^2j^2\bigg] \\ \label{helicityp}
  H_p &=& \int d^2r \big\{ab + \nonumber \\
    && ((1-\del)\eps)\left[(u+\del\eps~j)b+a(\omega-\del\eps\nabla^2 b)\right] + \nonumber \\
    && ((1-\del)\eps)^2\left[(u+\del\eps~j)(\omega-\del\eps\nabla^2 b)\right]\big\}  \\ \label{helicitye}
  H_e &=& \int d^2r \big\{ab - \nonumber \\
    && (\del\eps)\left[(u-(1-\del)\eps~j)b+a(\omega+(1-\del)\eps\nabla^2 b)\right] + \nonumber \\
    && (\del\eps)^2\left[(u-(1-\del)\eps~j)(\omega+(1-\del)\eps\nabla^2 b)\right]\big\}
\end{eqnarray}

In the present paper, we performed 2.5D EIHMHD~ simulations using a pseudo-spectral code, which yields exponentially fast numerical convergence and negligible numerical dissipation. The accuracy of the numerical scheme can be verified in part by looking at the behavior of the ideal invariants of the EIHMHD equations in time. The simulations reported here correspond to zero viscosity and resistivity, and the total energy is conserved by the numerical scheme with an error $\Delta E/E$ of less than $10^{-8}$. The ion and electron helicities were initially zero, and throughout their evolution differ from zero in less than $10^{-15}$. It is clear that numerical dissipation is reduced to round-off errors only. 

\section{Results}\label{results}

\subsection{Initial Conditions}\label{initcond}

Our initial condition to simulate a thin current sheets is given by (assuming periodic boundary conditions in a $2\pi\times2\pi$ box)
\begin{equation}\label{init-cond}
\textbf{B}(x,y,t=0) = \text{B}_0 \bigg[\tanh\left(\frac{y-\frac{3\pi}{2}}{2\pi\Delta}\right) - \tanh\left(\frac{y-\frac{\pi}{2}}{2\pi\Delta}\right) +1\bigg]\hat{\textbf{x}} 
\end{equation}
where, in normalized units, we have $\text{B}_0=1$ and $\Delta=0.02$. To drive reconnection, a monochromatic perturbation $\delta\textbf{B}=\boldsymbol\nabla\times[\hat{\textbf z}~\delta a(x,y)]$ with $\delta a(x,y) = a_0\cos(k_xx)$, $k_x = 1$ and an amplitude of $a_0=0.02\text{B}_0$ is added to the initial condition \eqref{init-cond}. We perform numerical simulations starting with a moderate spatial resolution of $512^2$ grid points, followed by progressively higher spatial resolutions of  $1024^2$, $1536^2$ and $2048^2$ grid points. For all these cases we use a Hall parameter $\varepsilon_{H}=0.1$ and a value of mass ratio $m_e/m_p = 0.015$, which corresponds to approximately 27 times the real electron mass. In addition, we made 3 runs with high spatial resolution ($1024^2$, $1536^2$ and $2048^2$ grid points) and a realistic ratio of electron to proton mass, i.e. $m_e/m_p = 1/1836$.

\begin{figure}
\centering
\includegraphics[width=.45\textwidth]{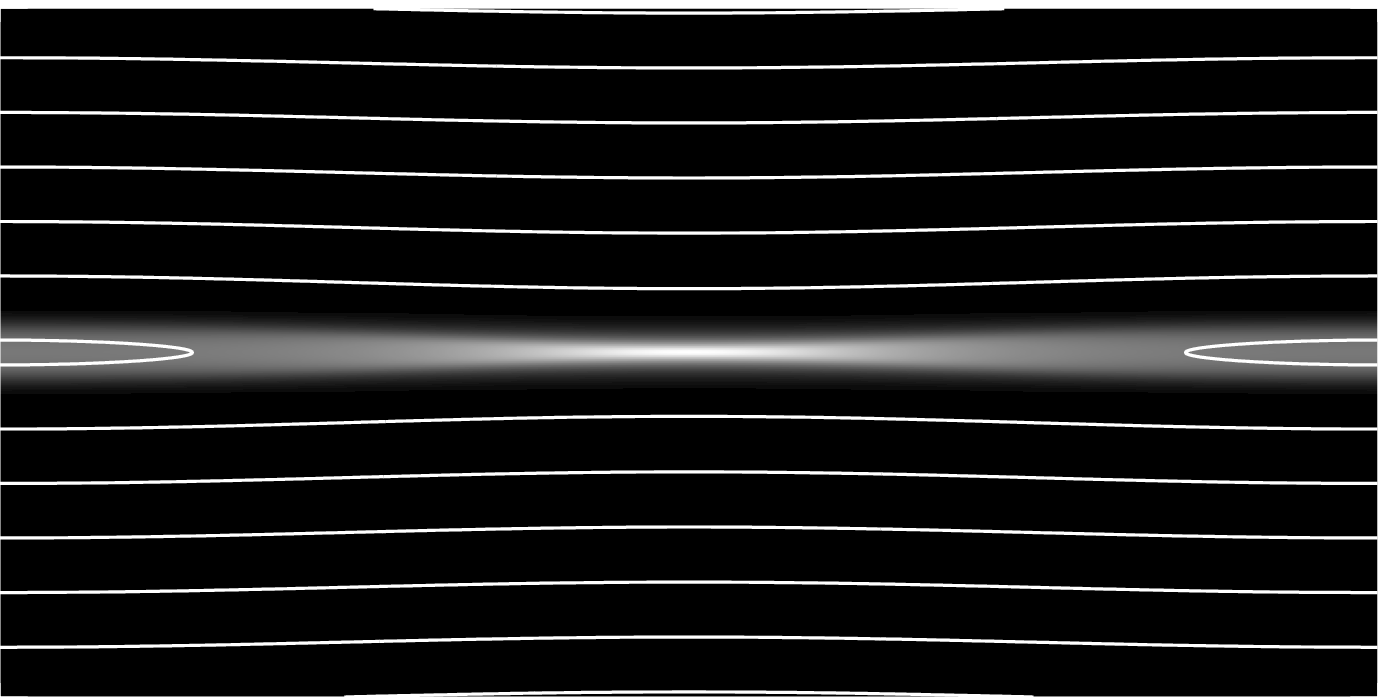} 
\\\vspace{.3cm}
\includegraphics[width=.45\textwidth]{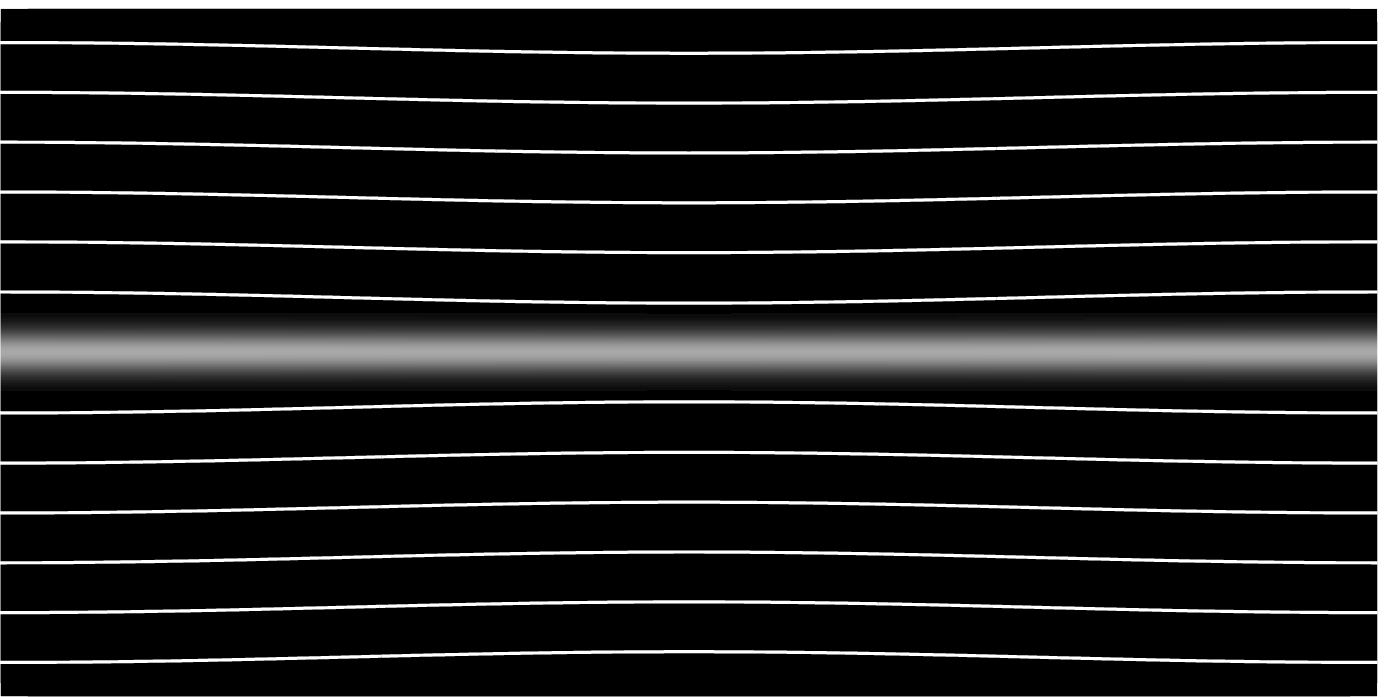}
\caption{The images (in grayscale) show the spatial distribution of current density $j(x,y)$ at $t=0.6$ for $\eps_H = 0.1$ and $m_e/m_p = 0.015$ (above) and $m_e/m_p = 0$ (below), for the lower half of the integration box (see Figure \ref{scheme}). Contours of $a(x,y)$ are superimposed (white lines).}\label{jj}
\end{figure}

In Figure \ref{jj} we show the set up of magnetic reconnection for a run of $1024^2$ grid points. Contours levels of magnetic flux $a(x,y)$ are in white lines, superimposed to the electric current density component along the $z$ direction, $j(x,y)$, at time $t=0.6$ (in grayscale). The panel above shows a EIHMHD run with $\eps_H = 0.1$ and $m_e/m_p = 0.015$, while the panel below shows a HMHD run with $\eps_H = 0.1$ and $m_e/m_p = 0$. The brightest regions correspond to the current sheets. We only show half a box of integration for each case, of size $2\pi\times\pi$.

\subsection{Topological change and spatial resolution}\label{topological}

As discussed in section \ref{intro}, magnetic reconnection is a local change in the magnetic field topology. One of the consequences of this topological change, is a transfer of free magnetic energy into kinetic energy of the plasma. In a fluid description, where resistivity and viscosity are set equal to zero, we expect that the break of the frozen-in condition is due to the presence of electron inertia. Therefore, we study the generation of magnetic reconnection in the following three models: MHD, HMHD, and EIHMHD, expecting that the only framework where reconnection is possible is EIHMHD.

\begin{figure}
\centering
\includegraphics[width=.4\textwidth]{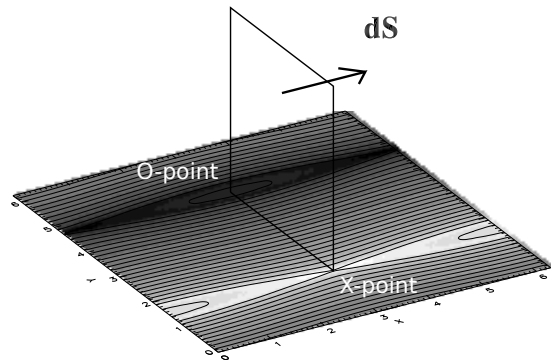}
\caption{Schematic configuration for the calculation of the reconnection rate. The horizontal plane shows the distribution of $j(x,y)$ for the full box, contour levels of $a(x,y)$ are superimposed.}\label{scheme}
\end{figure}

To quantitatively measure the efficiency of the magnetic reconnection process, the reconnection rate $r(t)$ is defined, which is the rate at which magnetic flux flows into the central neutral point (the X-point). Near the neutral point, magnetic flux enters due to a relatively slow plasma inflow and is expelled out at speeds of the order of the Alfv\'en speed. Figure \ref{scheme} shows the vertical surface used to integrate the magnetic flux $\phi(t)=\int\textbf{dS}\cdot\textbf{B}$, that extends from the O-point of one of the current sheets (shown in black, corresponding to negative values of $j(x,y)$), to the X-point of the other (shown in white). Both the O-point and the X-point are stagnation points of the flow. 

Using equation \eqref{Bmag} it is straightforward to show that
\begin{equation}
\phi(t) = \int\textbf{dS}\cdot\textbf{B} = a_{max} - a_{min}
\end{equation}
The reconnection rate $r(t)$ is the variation of this magnetic flux per unit time, i.e. $r(t)=d\phi(t)/dt$.

\begin{figure}
\centering
\includegraphics[width=.4\textwidth]{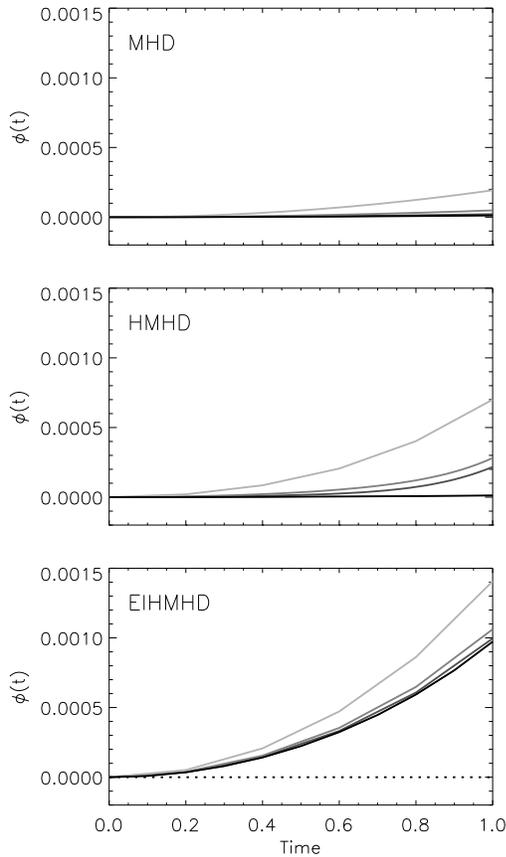}
\caption{Reconnected flux versus time. Each panel corresponds to a different case, as labelled. Different spatial resolutions: $512^2$, $1024^2$, $1536^2$, and $2048^2$ correspond to progressively darker traces.}\label{recfluxs}
\end{figure}

To test the accuracy of our results, we focused our attention on the spatial resolution of our simulations. For this purpose, we made different runs for several spatial resolutions, starting from the same initial condition as the one discussed in section \ref{idealinv}. More specifically, we performed 2.5D runs with the following numbers of grid points: $512^2$, $1024^2$, $1536^2$ and $2048^2$. For each spatial resolution, we calculated the reconnected flux as shown in Figure \ref{recfluxs}. As expected, for the ideal MHD and HMHD the curve for the reconnected flux converges to zero as the spatial resolution is increased (line color scale is obscured). Therefore, as the number of grid points increases, the reconnection rate approaches zero, both in the MHD and HMHD cases. In the case of EIHMHD, since we expect the electrons to break the frozen-in condition, the reconnected flux converges to a value different from zero, as the number of grid points increases. Therefore, we cla
 im that considering electron inertia is a necessary physical ingredient to start the reconnection process. Note that in our pseudo-spectral scheme, numerical dissipation is essentially zero (within round-off errors), as becomes apparent from the high degree of numerical conservation of the three ideal invariants (see also \citet{Br2013}).

\subsection{Magnetic Reconnected Flux and Rate}\label{recrate}

Within the framework of EIHMHD, we study the collisionless magnetic reconnection problem considering $\eps_H=0.1$ and a realistic value of the electron mass ($m_e/m_p=1/1836$). Using the same initial conditions described in section \ref{initcond}, we performed simulations with relatively high spatial resolution ($1024^2$, $1536^2$ and $2048^2$).
\begin{figure}
\centering
\includegraphics[width=.45\textwidth]{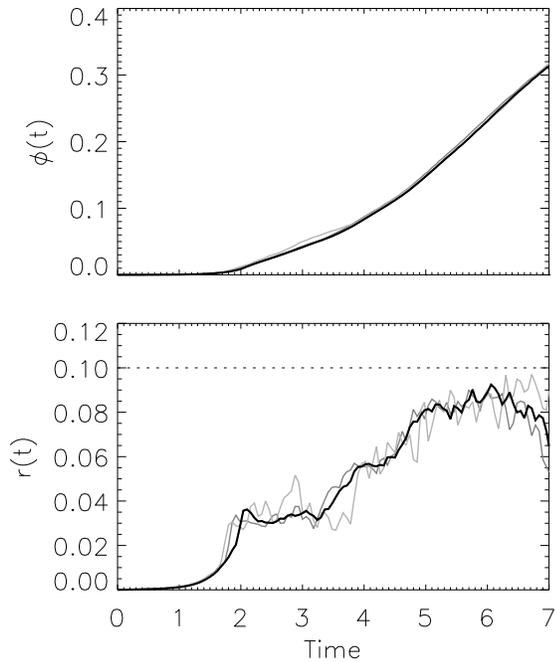}
\caption{Reconnected flux and reconnection rate as a function of time for $1024^2$ (light gray line), $1536^2$ (dark gray line) and $2048^2$ (black line) grid points. The three runs correspond to $\eps_H=0.1$ and a realistic mass ratio ($m_e/m_p=1/1836$).}
\label{real}
\end{figure}
Figure \ref{real} shows the reconnected flux and reconnection rate, for each spatial resolution, as a function of time. The reconnection rate $r(t)$ is calculated using second order finite central differences from the time series of the flux $\phi(t)$. As expected, we obtain essentially the same curve for the three spatial resolutions in agreement with the results shown in Figure \ref{recfluxs}. In particular, we get a maximum reconnection rate reaching values close to 0.1, which corresponds to inflow velocities approaching a fraction of the Alfv\'en speed. {This result is consistent with those reported in the literature, in particular with PIC simulation results \citep{Z2011,F2011} and the GEM Challenge \citep{Bi2001}. In particular, using a partially implicit PIC code, \citet{Z2011} found a reconnection rate approaching $0.1 v_A$.} It is worth mentioning that we obtained a reconnection rate comparable to the one reported by \citet{Bi2001}, because we used a similar set of initial conditions and 
parameter values. Nevertheless, the reconnection rate is expected to depend on the Hall parameter \citep{S2008,Bi1997,Sm2004,Mob2005,Mo2006} as well as on the amplitude of fluctuations $\delta$B\citep{Sm2004}, even though a systematic study of the reconnection rate as a function of these parameters is beyond the scope of the present study.

{Finally, we compared the reconnected flux and reconnection rate for the same initial conditions and different electron to proton mass ratios. In particular, we compared the results for $m_e/m_p=0.015$ and $m_e/m_p=1/1836$. We obtain the same trend for both the reconnected flux and reconnection rate. In agreement with PIC simulations \citep{Z2011} and resistive HMHD simulations \citep{Bi2001}, we find that the reconnection rate is insensitive to the electron to proton mass ratio.}

\section{Conclusions}\label{conclus}

We presented a fully two-fluid model for a completely ionized hydrogen plasma, retaining the Hall current and electron inertia. In the incompressible limit, we verified the existence of the linear modes of the model, i.e. the ion-cyclotron and the whistler branches. As showed, these two branches converge to the proton and electron cyclotron frequencies in the wavenumber $k\rightarrow\infty$ limit. Numerically, we confirm the conservation of the three ideal invariants of the model with a high degree of accuracy of $\sim10^{-8}-10^{-12}$. It is worth mentioning that in the limit of zero electron inertia (i.e. $m_{e}\rightarrow0$) we recover the HMHD model, with their corresponding linear modes and ideal invariants \citep{G2008}.

Our results show that we are able to obtain magnetic reconnection, only when the effects of electron inertia are retained, since our scheme is free from physical or numerical resistivity. {Even though the fact that electron inertia enables magnetic reconnection is well known, to the extent of our knowledge this is the first time that this feature is confirmed  with results from a non-dissipative fluid simulation.} In particular, for the case of ideal MHD and HMHD, we show that it is not possible to have magnetic reconnection without dissipation effects. In other words, we find that within the framework of the present model, finite electron inertia is a necessary physical ingredient to drive the reconnection process, even though the reconnection rate is largely independent of the numerical value of the mass ratio $m_e/m_p$. Moreover, for high spatial resolution simulations we find a reconnection rate that is quantitatively compatible with the one found by \citet{Bi2001}, when we use parameter values 
and initial conditions similar to theirs. Note however, that the reconnection rate might still depend on the value of the Hall parameter $\eps_H$ or on the level of fluctuations $\delta$B.

\section*{Acknowledgments}

The authors Nahuel Andrés and Luis Martin would like to acknowledge interesting discussions and advice by Vanesa Monserrat that helped us to make this work possible.

We acknowledge support from the following grants PIP 11220090100825, UBACyT 20020110200359, 20020100100315, and PICT 2011-1529, 2011-1626 and 2011-0454.

% \bibliographystyle{apsrev4-1}
% \bibliography{cites}

\end{document}